\documentclass{WileyMSP-template}

\begin{document}

\pagestyle{fancy}
\rhead{\includegraphics[width=2.5cm]{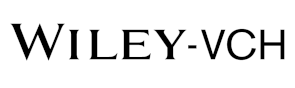}}

\title{Digital-Analog Quantum Machine Learning}

\maketitle


\author{Lucas Lamata}*


\dedication{Dedicated to the founders of Quantum Mechanics}

\begin{affiliations}
L. Lamata\\
Departamento de F\'isica At\'omica, Molecular y Nuclear, Facultad de F\'isica, Universidad de Sevilla, Apartado 1065, E-41080 Sevilla, Spain\\
Email Address: llamata@us.es

\end{affiliations}



\begin{abstract}

Machine Learning algorithms are extensively used in an increasing number of systems, applications, technologies, and products, both in industry and in society as a whole. They enable computing devices to learn from previous experience and therefore improve their performance in a certain context or environment. In this way, many useful possibilities have been made accesible. However, dealing with an increasing amount of data poses difficulties for classical devices. Quantum systems may offer a way forward, possibly enabling to scale up machine learning calculations in certain contexts. On the other hand, quantum systems themselves are also hard to scale up, due to decoherence and the fragility of quantum superpositions. In the short and mid term, it has been evidenced that a quantum paradigm that combines evolution under large analog blocks with discrete quantum gates, may be fruitful to achieve new knowledge of classical and quantum systems with no need of having a fault-tolerant quantum computer. In this Perspective, we review some recent works that employ this digital-analog quantum paradigm to carry out efficient machine learning calculations with current quantum devices.

\end{abstract}


\section{Quantum Machine Learning}

Machine learning is a knowledge field, a tool, a computing paradigm, a technology, which is significantly impacting society at large, by enabling a plethora of possibilities, as well as challenges, as with any new and highly powerful technology. However, its full deployment is hampered by the fact that an increasing amount of data is hard to process with current classical computers, and consumes a significant amount of energy. A possible way forward to scale up machine learning calculations would be via the use of quantum devices employing genuine quantum properties, such as superposition and entanglement. In this way, it may be possible to take advantage of the speedup of quantum computers to carry out machine learning calculations in a more efficient way, at least in some instances. Even though the field of quantum machine learning, which connects machine learning calculations to a quantum hardware seeking for these advantages, is not absent of controversy and speculation, as it is really hard to prove speedup with respect to the best classical machine learning algorithms, this is a field that is rising much expectation and hope that with NISQ devices one may beat classical computers for useful tasks. For a recent review of the field, see Ref. [1].

\section{Digital-Analog Quantum Paradigm}

Full-fledged, scalable, digital quantum computers are hard to achieve, as they normally require millions of physical qubits to encode fault-tolerance and error-correction protocols. On the other hand, analog quantum simulators are typically scalable but limited in the amount of problems they are able to solve, or reproduce in a quantum simulation. A possible combination of both approaches has emerged in the past few years as a strategy for the short and mid term, which may enable NISQ quantum devices to achieve useful tasks, for instance in learning new properties of quantum systems. The focus on the field of digital-analog quantum protocols has been mostly on quantum simulations in the past few years, with theory proposals and  experiments in a variety of quantum platforms. However, most recently several works that aim at carrying out machine learning calculations via digital-analog quantum protocols have appeared in the literature. For an early review on the field of digital-analog quantum simulations see Ref. [2].

In Fig. 1 I plot a scheme of a generic {\it extreme} digital-analog quantum protocol (EDAQP). This consists of a combination of the largest analog blocks possible (a set of global native interactions, namely, global unitary gates on all qubits based on the quantum platform native Hamiltonian, $U_1(t_1)$, $U_2(t_2)$,...,$U_k(t_k)$,..., and the smallest digital gates possible, i.e., single-qubit gates, $R_{i,j}$, acting on qubit j at step i, which can be done with the best fidelities in most quantum platforms. 

 \begin{figure}
  \includegraphics[width=\linewidth]{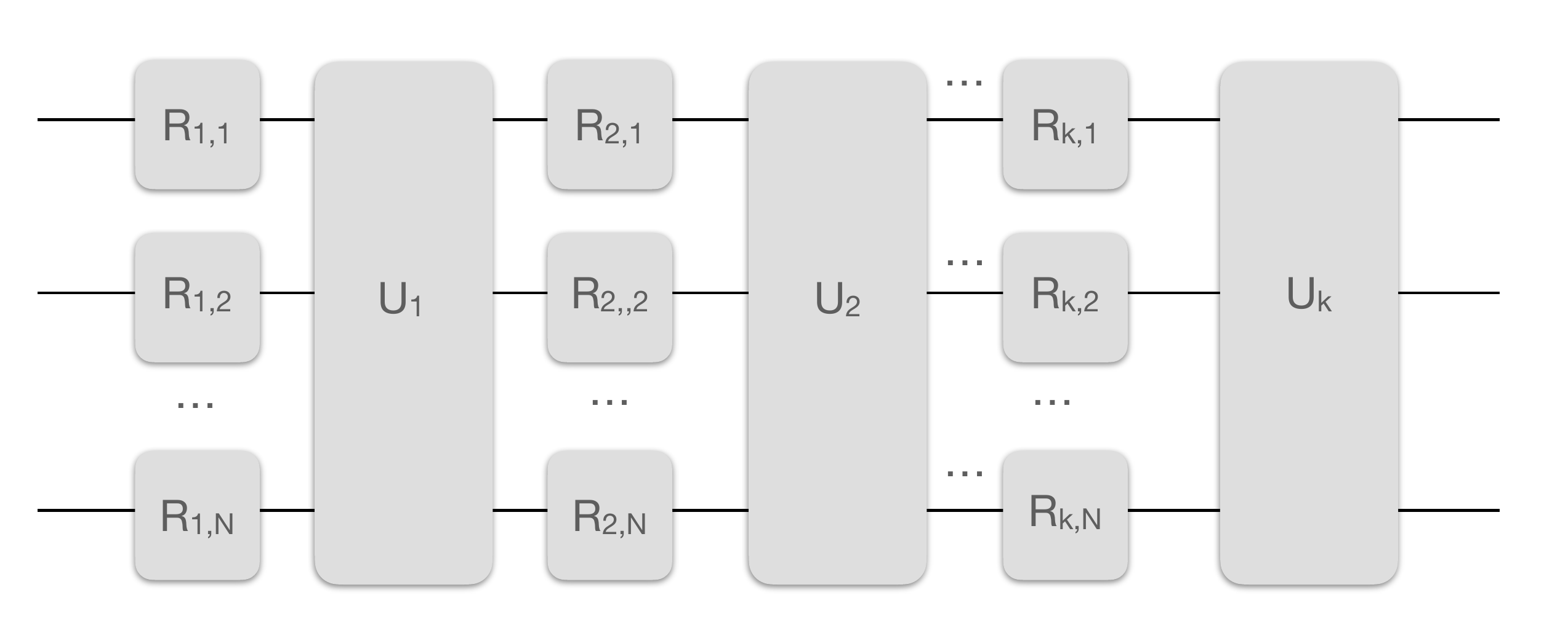}
  \caption{scheme of a generic {\it extreme} digital-analog quantum protocol (EDAQP). This consists of a combination of the largest analog blocks possible, i.e., a set of global native interactions, namely, global unitary gates on all qubits based on the quantum platform native Hamiltonian, $U_1(t_1)$, $U_2(t_2)$,...,$U_k(t_k)$,..., and the smallest digital gates possible, i.e., single-qubit gates, which can be done with the best fidelities in most quantum platforms.}
\end{figure}

\section{Digital-Analog Quantum Machine Learning}

In this Perspective, I give a non-exhaustive overview of the field of quantum machine learning, when digital-analog quantum protocols are employed for its deployment. I propose to use the name ``Digital-analog quantum machine learning'' (DAQML) for this research avenue. I will cite several papers in this area as well as briefly describe each of them.

In Ref. [3], the authors propose to use a digital-analog quantum protocol to implement a variational quantum eigensolver for estimating ground state energies of molecules. A subsequent work [4], analyzed the deployment of a quantum genetic algorithm via the Qadence digital-analog coding language for Rydberg atom arrays [5], to address also calculations of ground state energies of molecules, in a scalable way. This instance of digital-analog quantum protocol was of the EDAQP as shown in Fig. 1. Ref. [6] studies the use of a digital-analog quantum approximate optimization algorithm for solving a series of tasks with NISQ devices. In Ref. [7], the authors propose to use a digital-analog quantum paradigm for quantum kernel implementation in the context of quantum machine learning. In Ref. [8], a quantum Fourier transform implementation is proposed via the use of digital-analog quantum protocols, showing possible gains in resources with respect to purely digital ones, and enabling in this way the use of this primitive in a variety of protocols, which may also include quantum machine learning ones, in particular. This is for instance analyzed in Ref. [9] with respect to the Harrow-Hassidim-Lloyd [HHL] protocol. In Ref. [10], the authors propose digital-analog quantum learning algorithms for Rydberg atom systems, which they claim can combine the usefulness of quantum machine learning protocols in the near term, with the efficient scalability that is recently being achieved with Rydberg atoms.

Most previous works, which are first instances of the combination of quantum machine learning algorithms with the digital-analog quantum paradigm, show evidence that this combination can be fruitful, and perhaps, by enabling a better scalability of quantum machine learning algorithms, may provide a quantum advantage with respect to classical machine learning calculations.

\section{Implementations of the Digital-Analog Quantum Paradigm}

Several experiments in quantum platforms have reached a significantly large amount of qubits and/or Hilbert space dimension, by combining state-of-the-art developments in these platforms, and the use of digital-analog or similar quantum techniques. Some of the quantum platforms employed are trapped ions, superconducting circuits, and Rydberg atoms. This kind of approach has been considered in parallel by several groups, which employ large analog blocks combined with discrete gates, although some of these techniques are also referred to in the literature as ``programmable analog quantum simulations''. For a recent review that describes analog, digital, and digital-analog quantum paradigms, in the context of quantum simulation experiments, see Ref. [11].  Even though these pioneering experiments have mainly been focused on quantum simulations, similar setups may address quantum machine learning tasks as the ones described above, with possible gains with respect to classical computers at least in some favorable situations [See, e.g., work done in startups PASQAL [12] and QUERA [13] in this sense].

\section{Outlook}

The previous account of recent literature combining digital-analog quantum protocols with quantum machine learning algorithms, which I name ``Digital-Analog Quantum Machine Learning'' (DAQML), is just an appetizer for what may come in the future. With the increasing efficiency of quantum platforms such as trapped ions, superconducting circuits, and Rydberg atoms, in achieving devices with tens, or even, hundreds of quantum bits, the possibility to combine large analog blocks with digital steps, in the paradigm that we created more than 10 years ago in the context of quantum simulations, may represent a significant step forward in the field of quantum technologies. Perhaps one of the first approaches that could possibly impact the industry for carrying out useful tasks might be the one I describe here, of combining our digital-analog quantum paradigm with the fledgling field of quantum machine learning. This could, in turn, impact subsequently society at large, as well as the scientific enterprise as well, via enhancing scientific discovery, as this very new journal, Advanced Intelligent Discovery, aims to achieve.

\medskip
\textbf{Acknowledgements} \par 

I acknowledge the support from grants PID2022-136228NB-C21 and PID2022-136228NB-C22 funded by
MCIN/AEI/10.13039/50110001103 and “ERDF A way
of making Europe”. This work has also been financially
supported by the Ministry for Digital Transformation and of Civil Service of the Spanish Government
through the QUANTUM ENIA project call - Quantum
Spain project, and by the European Union through
the Recovery, Transformation and Resilience Plan -
NextGenerationEU within the framework of the “Digital
Spain 2026 Agenda”.

\medskip

%

\textbf{References}\\

1 Yunfei Wang and Junyu Liu, A comprehensive review of quantum machine learning: from NISQ to fault tolerance, Rep. Prog. Phys. 87 116402 (2024).\\

2 L Lamata, A Parra-Rodriguez, M Sanz, and E Solano, Digital-Analog Quantum Simulations with Superconducting Circuits, Adv. Phys.: X 3 (1), 1457981 (2018).

3 Antoine Michel, Sebastian Grijalva, Loïc Henriet, Christophe Domain, and Antoine Browaeys, Blueprint for a digital-analog variational quantum eigensolver using Rydberg atom arrays, Phys. Rev. A 107, 042602 (2023).\\

4 Aleix Llenas and Lucas Lamata, Digital-analog quantum genetic algorithm using Rydberg-atom arrays, Phys. Rev. A 110, 042603 (2024).\\

5 D. Seitz, N. Heim, J. P. Moutinho, R. Guichard, V. Abramavicius, A. Wennersteen, G.-J. Both, A. Quelle, C. de Groot, G. V. Velikova, V. E. Elfving, and M. Dagrada, Qadence:
a differentiable interface for digital-analog programs, arXiv:2401.09915 (2024).

6 David Headley, Thorge M\"uller, Ana Martin, Enrique Solano, Mikel Sanz, and Frank K. Wilhelm, Approximating the quantum approximate optimization algorithm with digital-analog interactions, Phys. Rev. A 106, 042446 (2022).

7 Anton Simen, Carlos Flores-Garrigos, Narendra N. Hegade, Iraitz Montalban, Yolanda Vives-Gilabert, Eric Michon, Qi Zhang, Enrique Solano, and Jos\'e D. Mart\'in-Guerrero, Digital-analog quantum convolutional neural networks for image classification, arXiv:2405.00548 (2024).

8 Ana Martin, Lucas Lamata, Enrique Solano, and Mikel Sanz, Digital-analog quantum algorithm for the quantum Fourier transform, Phys. Rev. Research 2, 013012 (2020).

9 Ana Martin, Ruben Ibarrondo, and Mikel Sanz, Digital-Analog Co-Design of the Harrow-Hassidim-Lloyd Algorithm, Phys. Rev. Applied 19, 064056 (2023).

10 Jonathan Z. Lu, Lucy Jiao, Kristina Wolinski, Milan Kornjaca, Hong-Ye Hu, Sergio Cantu, Fangli Liu, Susanne F. Yelin, and Sheng-Tao Wang, Digital-analog quantum learning on Rydberg atom arrays, arXiv:2401.02940 (2024).

11 Andrew J. Daley, Immanuel Bloch, Christian Kokail, Stuart Flannigan, Natalie Pearson, Matthias Troyer and Peter Zoller, Practical quantum advantage in quantum simulation, Nature 607, 667 (2022).

12 www.pasqal.com

13 www.quera.com

\end{document}